\documentclass[12pt]{article}
\begin{document}
\title{A Non-commutative Version of\\the Fundamental Theorem
of Asset Pricing}
\author{Zeqian Chen\\{\small Wuhan Institute of Physics and
Mathematics, Chinese Academy of Sciences}\\
{\small 30 West District, Xiao-Hong Mountain, Wuhan}\\{\small
P.O.Box 71010, Wuhan 430071}\\{\small E-mail: zqchen@wipm.ac.cn}}
\date{}
\maketitle

{\it Abstract}.~~In this note, a non-commutative analogue of the
fundamental theorem of asset pricing in mathematical finance is
proved.

\

\section*{1.~Introduction}

In retrospect, the field of mathematical finance has undergone a
remarkable development since the seminal papers by F.Black and
M.Scholes [2] and R.Merton [15], in which the famous
``Black-Scholes Option Pricing Formula" was derived. The idea of
developing a ``formula" for the price of an option actually goes
back as far as 1900, when L.Bachelier wrote a thesis with the
title ``Th\'{e}orie de la sp\'{e}culation" [1]. It was Bachelier
who firstly had the innovative idea of using a {\it stochastic
process} as a model for the price evolution of a stock. For a
stochastic process $(S_t)_{0 \leq t \leq T}$ he made a natural and
far-reaching choice being the first to give a mathematical
definition of Brownian motion, which in the present context is
interpreted as follows: $S_0$ is today's (known) price of a stock
(say a share of company XYZ to fix ideas) while for the time $t >
0$ the price $S_t$ is a normally distributed random variable.

The basic problem of Bachelier, as well as of modern Mathematical
Finance in general, is that of assigning a price to a contingent
claim. Bachelier used the equilibrium argument. It was the merit
of Black and Scholes [2] and Merton [15] to have replaced this
argument by a so-called ``no-arbitrage" argument, which is of
central importance to the entire theory. Roughly speaking, an
arbitrage is a riskless way of making a profit with zero net
investment. An economically very reasonable assumption on a
financial market consists of requiring that there are no arbitrage
opportunities. The remarkable fact is that this simple and
primitive ``principle of no arbitrage" allows already to determine
a unique option price in the Black-Scholes model. This is the
theme of the so-called {\it fundamental theorem of asset pricing}
which states briefly that a process $S= (S_t)$ does not allow
arbitrage opportunities if and only if there is an equivalent
probability measure under which $S$ is a martingale.

The history of the fundamental asset pricing theorem goes back to
the seminal work of Harrison, Kreps and Pliska ([11, 12, 14]).
After their pioneering work many authors made contributions to
gradually improve the understanding about this fundamental
theorem, e.g., Duffie and Huang [10], Stricker [21], Dalang,
Morton, and Willinger [6], and Delbaen and Schachermayer [7] etc.
In [8] this theorem was proved to hold true for very general
(commutative) stochastic processes.

In this note we deal with this issue in the non-commutative (=
quantum) setting. After having formalized the notations of
(quantum) arbitrage and quantum trading strategies, we shall prove
a non-commutative analogue of the fundamental theorem of asset
pricing. As shown in [4], there are several reasons why quantizing
mathematical finance may be interesting. In particular, classical
mathematical finance theory is a well established discipline of
applied mathematics (see [9, 20] and references therein) which has
found numerous applications in financial markets (see for example
[13, 16]). Since it is based on probability to a large extend,
there is a fundamental interest in generalizing this theory to the
domain of quantum probabilities. Indeed, recently non-commutative
(= quantum) probability theory has developed considerably. In
particular, all sorts of non-commutative analogues of Brownian
motion and martingales have been studied. We refer to [17] and
references therein. Moreover, it has recently been shown that the
quantum version of financial markets is maybe much more suited to
real-world financial markets rather than the classical one,
because the quantum binomial model ceases to pose the paradox
which appears in the classical model of the binomial market, see
[3, 5] for details.

\section*{2.~Notational preliminaries and the main result}

Throughout this note we shall denote by $({\cal A}, \tau)$ a
$W^*$-non-commutative probability space, namely, ${\cal A}$ is a
finite von Neumann algebra, and $\tau$ is a faithful normal
tracial state on ${\cal A}.$ (See [18, 23] for details on von
Neumann algebras.) We shall denote by $L^p ({\cal A}, \tau)$ or
simply $L^p ({\cal A})$ the non-commutative $L^p$-spaces. Note
that if $p =\infty, L^p ({\cal A})$ is just ${\cal A}$ itself with
the algebra norm; also recall that the norm in $L^p ({\cal A})$
($1 \leq p < \infty$) is defined as$$\| a \|_p = \tau [ |a|^p
]^{\frac{1}{p}},$$ where $|a| = (a^* a)^{1/2}$ is the usual
absolute value of $a.$ We shall assume that ${\cal A}$ is
filtered, so that there exists a family $( {\cal A}_t )_{t \in
{\bf R}_+}$ of unital weakly closed $*$- subalgebras of ${\cal
A},$ such that ${\cal A}_s \subset {\cal A}_t$ for all $s,t$ with
$s \leq t,$ and ${\cal A}_0 = {\bf C} I, ~I$ denoting the unit
element in ${\cal A}.$ Since the state $\tau$ is tracial, for any
unital weakly closed $*$-subalgebra ${\cal B}$ of ${\cal A},$
there exists a unique conditional expectation onto ${\cal B}.$ We
shall denote by $E_{\tau} [. | {\cal B}]$ this conditional
expectation. Recall that it extends to a contraction on all
$L^p$-spaces for $1 \leq p \leq \infty.$ A map $t \to M_t$ from
$[0, +\infty )$ to $L^p ( {\cal A}, \tau)$ will be called a
martingale with respect to the filtration $( {\cal A}_t )_{t \in
{\bf R}_+}$ if for every $s \leq t$ one has that $E_{\tau} [ M_t |
{\cal A}_s ] = M_s.$

However, even for a state $\sigma$ in a finite dimensional von
Neumann algebra ${\cal A}$ the conditional expectation operator
$E_{\sigma} [.|{\cal B}]$ of a $*$-subalgebra ${\cal B}$ of ${\cal
A}$ does not need to exist in general (for details see [22]). Thus
we cannot define a martingale under $\sigma$ as in the case of the
tracial states or the commutative setting. It seems to us that one
needs to generalize the definition of martingales in the
non-commutative setting as following:

\

{\it Definition 1}.~~Given any fixed state $\sigma$ on ${\cal A}.$
A family $\{ M_t \}_{t \geq 0}$ in ${\cal A}$ is said to be a
$(non-commutative)$ martingale with respect to $({\cal A}, ({\cal
A}_t )_{t\geq 0}, \sigma)$ if it is adapted to $({\cal A}_t
)_{t\geq 0}$ and for every $0\leq s \leq t,$ $$\sigma (a M_t a^* )
= \sigma (a M_s a^* ),$$for all $a \in {\cal A}_s.$

\

Clearly, when $\sigma$ is a normal tracial state the above
definition coincides to the usual definition of the
non-commutative martingales. In the sequel we understand the
non-commutative martingales in this sense. We would like to point
out that those martingales in the above sense are suitable in the
so-called quantum finance, for details see [4].

Together with $({\cal A}, \tau)$ we shall also consider the
opposite algebra ${\cal A}^{op},$ with the trace $\tau^{op},$
namely $\tau = \tau^{op}$ as a linear map on ${\cal A},$ but the
notation is meant to stress the algebra structure we are using.
The spaces ${\cal A}$ and ${\cal A} \otimes {\cal A}$ have natural
${\cal A}-{\cal A}$ bimodule structures given by multiplication on
the right and on the left, namely $a.u.b = a u b$ and $a. ( u
\otimes v). b = a u \otimes v b,$ or equivalently they have a left
${\cal A} \otimes {\cal A}^{op}$-module structure. We shall denote
by $\sharp$ these actions, namely one has $(a \otimes b) \sharp u
= a u b$ and $(a \otimes b) \sharp (u \otimes v) = (a u) \otimes
(v b).$ The map $\tau \otimes \tau^{op}$ defines a tracial state
on the $*$-algebra ${\cal A} \otimes {\cal A}^{op},$ and we shall
denote by $L^p (\tau \otimes \tau^{op} )$ the corresponding $L^p$-
spaces, thus $L^{\infty} (\tau \otimes \tau^{op})$ is the von
Neumann algebra tensor product of ${\cal A}$ and ${\cal A}^{op}.$

A simple biprocess is a piecewise constant map $t \to H_t$ from
${\bf R}_+$ into the algebraic tensor product ${\cal A} \otimes
{\cal A}^{op},$ such that $H_t =0$ for $t$ large enough. It is
called to be adapted if one has $H_t \in {\cal A}_t \otimes {\cal
A}_t$ for all $t \geq 0.$ In this case, it is clear that one can
choose a decomposition\begin{equation} H_t = \sum^n_{j=1} A_{j,t}
\otimes B_{j,t}\end{equation}such that there exist times $0=t_0
\leq t_1 \leq... \leq t_m$ with $A_{j,t} = A_{j,t_k},~B_{j,t} =
B_{j,t_k} \in {\cal A}_{t_k}$ for $t \in [ t_k, t_{k+1}), A_{j,t}
= B_{j,t} = 0$ for all $t \geq t_m$ (in the sequel we shall always
assume that the decompositions we choose satisfy such properties).

In the sequel we always assume that $X = (X_t)_{t \geq 0}$ is a
self-adjoint stochastic process adapted to the filtered space
$({\cal A}, ({\cal A})_{t\geq 0}),$ i.e., for every $t \geq 0, X_t
\in {\cal A}_t$ and $X^*_t = X_t.$

\

{\it Definition 2}.~~Let $H$ be a simple adapted biprocess with a
decomposition as above, then the stochastic integral of $H$ with
respect to $X = (X_t)_{t \geq 0}$ is\begin{equation}
\int^{\infty}_0 H_s \sharp d X_s = \sum^{m-1}_{k=0} H_{t_k} \sharp
( X_{t_{k+1}} - X_{t_k} ) = \sum^{m-1}_{k=0} \sum^{n_k}_{j=1}
A_{j,t_k} ( X_{t_{k+1}} - X_{t_k} ) B_{j,t_k}.\end{equation} This
is clearly independent of the decomposition chosen.

For a simple adapted biprocess $H,$ and $s < t,$ we shall denote
$H^{(s,t)}$ the stopped simple adapted biprocess given by
$H^{(s,t)}_r = H_r$ for $s \leq r <t$ and $H^{(s,t)}_r =0$ for $r
< s$ or $r \geq t.$ Then we define$$ \int^t_s H_r \sharp d X_r =
\int^{\infty}_0 H^{(s,t)}_r \sharp d X_r.$$We shall write $(H
\sharp X )_t = \int^t_0 H_r \sharp d X_r.$

\

{\it Remark 1}.~~The space of adapted simple biprocesses has an
antilinear involution, coming from the antilinear involution on
${\cal A} \otimes {\cal A}$$$( \sum A_j \otimes B_j)^* = \sum
B^*_j \otimes A^*_j.$$The adjoint of the stochastic integral is
again a stochastic integral, namely with the adjoint of a
biprocess as above, one has that$$(\int^{\infty}_0 H_t \sharp d
X_t )^* = \int^{\infty}_0 H^*_t \sharp d X_t.$$

\

{\it Definition 3}.~~${\cal H}$ denotes the set of simple {\it
quantum trading strategies} for $X = (X_t)_{t \geq 0}.$ An element
$H = (H_t)_{t \geq 0} \in {\cal H}$ is a simple biprocess of the
form $$H_t = \sum {\alpha}_j a_j \otimes a^*_j,$$with $a_j \in
{\cal A}_t,$ where $\alpha_j$ are all real numbers.

\

{\it Remark 2}.~~Evidently, $\int^{\infty}_0 H_t \sharp d X_t$ is
self-adjoint provided $H \in {\cal H}.$

\

We define $K^s$ the set of all self-adjoint elements of form $(H
\circ X)_{\infty},$ where $H \in {\cal H},$ and $C^s$ the convex
cone of self-adjoint elements $a$ in ${\cal A}$ with the property
that $a \leq b$ for some $b \in K^s.$ We denote by $\bar{C}^*$ the
closure of $C^s$ with respect to the weak-star topology $\sigma
({\cal A}, {\cal A}_*)$ of ${\cal A},$ where ${\cal A}_*$ is the
predual space of ${\cal A}.$ It is well known that $${\cal A}_* =
L^1 ({\cal A}, \tau)$$ via the correspondence that $$b \to \tau[a
b],~~b \in L^1 ({\cal A}, \tau),$$for each $a \in {\cal A}.$

\

{\it Definition 4} (e.g., [14]).~~We say that $X = (X_t)_{t \geq
0}$ satisfies the condition of no free lunch (NFL)
if\begin{equation}\bar{C}^* \cap {\cal A}_+ = \{ 0
\}.\end{equation}

\

{\it Definition 5}.~~A normal state $\sigma$ on ${\cal A}$ is
called a martingale state of $X = (X_t)_{t \geq 0},$ if $X =
(X_t)_{t \geq 0}$ is a martingale on $({\cal A}, ({\cal A})_{t\geq
0}, \sigma ).$

We denote by $M_f(X)$ the family of all such faithful normal
states, and say that $X = (X_t)_{t \geq 0}$ satisfies the
condition of the existence of a faithful martingale state (EMS) if
$M_f(X) \not= \emptyset.$

\

As following is a non-commutative analogue of the fundamental
theorem of asset pricing in mathematical finance:

\

{\bf Theorem}.~~{\it A non-commutative self-adjoint stochastic
process $X = (X_t)_{t \geq 0}$ satisfies the condition of no free
lunch $($NFL$)$ if and only if the condition $($EMS$)$ of the
existence of a faithful martingale state is satisfied.}

\

{\it Remark 3}.~~In [4] the author has proved a special case of
the above theorem on finite dimensional von Neumann algebras,
whose proof is different from that presented here. By using this
theorem we present a quantum version of the classical asset
pricing theory of multi-period financial markets based on finite
dimensional quantum probability spaces.

\section*{3.~Proofs}

{\bf Lemma 1}.~~{\it Let $H$ be in ${\cal H}$ and let $\sigma$ be
a state on ${\cal A}.$ If $X = (X_t)_{t \geq 0}$ is a martingale
under $\sigma,$ then $t \to (H \sharp X )_t$ is also a martingale
under $\sigma.$}

\

{\it Proof.}~~Let $H_t = a \otimes a^* 1_{[t_1, t_2)} (t)$ where
$a \in {\cal A}_{t_1}.$ Let $s \leq t$ and $y \in {\cal A}_s.$ We
have to prove that$$ \sigma[y \int^t_s H_r \sharp d X_r y^* ]
=0.$$One has that$$\int^t_s H_r \sharp d X_r = a (X_{\min (\max
(t, t_1), t_2)} - X_{\max (\min (s, t_2), t_1)} )a^*.$$ Since $X =
(X_t)_{t \geq 0}$ is a martingale, we get the result. The general
case follows since linear combinations of martingales are
martingales.

\

{\bf Lemma 2}.~~{\it Let $\sigma$ be a state on ${\cal A}.$ Then,
$X = (X_t)_{t \geq 0}$ is a martingale under $\sigma$ if and only
if$$\sigma [ (H \sharp X )_{\infty} ] = 0,$$for every $H \in {\cal
H}.$}

\

{\it Proof.}~~Suppose that $X = (X_t)_{t \geq 0}$ is a martingale.
By Lemma 1 one concludes that$$\sigma [ (H \sharp X )_{\infty} ] =
\sigma [ (H \sharp X )_0 ] =0,$$for every $H \in {\cal H}.$

Conversely, let $s \leq t$ and $y \in {\cal A}_s.$ Set $H_r = y
\otimes y^* 1_{[s, t)} (r).$ Then $$(H \sharp X )_{\infty} = y
(X_t - X_s) y^*,$$and hence $\sigma [ y ( X_t - X_s ) y^* ] =0.$
The proof is complete.

\

{\it Proof of the Theorem}.~~(EMS) $\Longrightarrow$ (NFL):~By
Lemma 2 we have that $\sigma (c) \leq 0$ for each $\sigma \in M_f
(X)$ and $c \in C^s,$ and this inequality also extends to the
weak-star closure $\bar{C}^*.$ However, if (EMS) would hold and
(NFL) were violated, there would exist a $\sigma \in M_f(X)$ and
$c \in \bar{C}^*, c > 0,$ whence $\sigma (c) > 0$ since $\sigma$
is faithful, a contradiction.

(NFL) $\Longrightarrow$ (EMS):~We claim that, for fixed $a_0 \in
{\cal A}, a_0 > 0,$ there is $b \in L^1 ({\cal A})$ which defines
a positive linear functional $\tau_b$ on ${\cal A}$ via $$\tau_b
(a) = \tau [a b], ~~a \in {\cal A},$$such that $\tau_b$ is less or
equal to zero on $\bar{C}^*,$ and $\tau_b (a_0) > 0.$ To see this,
apply the separation theorem (e.g., [19, Theorem II.9.2]) to the
$\sigma ({\cal A}, {\cal A}_*)$-closed convex set $\bar{C}^*$ and
the compact set $\{ a_0 \}$ to find a $b \in L^1 ({\cal A})$ and
$\alpha < \beta$ such that $\tau_b [ c] \leq \alpha$ for all $c
\in \bar{C}^*$ and $\tau_b (a_0 ) > \beta.$ Since $0 \in C^s$ we
concludes that $\alpha \geq 0.$ As $\bar{C}^*$ is a cone, we have
that $\tau_b$ is zero or negative on $\bar{C}^*$ and, in
particular, nonnegative on ${\cal A}_+.$ Noting that $\beta > 0$
we have proved the claim.

Denote by ${\cal B}$ the set of all $b \in L^1 ({\cal A})$ so that $\tau_b$ is
a positive linear functional on ${\cal A}$ which is less or equal
to zero on $\bar{C}^*.$
Clearly $0 \in {\cal B}$ and hence ${\cal B}$ is nonempty.

Let ${\cal S}$ be the set of all supports $s(\tau_b)$ of $\tau_b,
b \in {\cal B}.$ Note
that ${\cal S}$ is a $\sigma$-lattice in the usual order, as for a
sequence $b_n \in {\cal B},$ we may find strictly positive scalars $\alpha_n$
such that $\sum_n \alpha_n b_n \in {\cal B}.$ Hence there is $
b_0 \in {\cal B} $ such that$$s(\tau_{b_0} ) = \sup \{
s(\tau_b):~ b \in {\cal B} \}.$$

We now claim that $s(\tau_{b_0} ) = 1,$ which readily shows that $\tau_{b_0}$
is faithful. Indeed, if $s(\tau_{b_0}) < 1,$ then we could apply the above
claim to $1 - s(\tau_{b_0})$ to find $b_1 \in {\cal B}$ with$$\tau [b_1
(1- s(\tau_{b_0}))] > 0.$$Hence, $b_0 + b_1$ would be an element
of ${\cal B}$ whose support is bigger than $s(\tau_{b_0}),$ a contradiction.

Normalize $\tau_{b_0}$ so that $\tau_{b_0} [1] =1,$ we concludes from Lemma 2
that $\sigma = \tau_{b_0}$ is a martingale state for $X$ and thus, $M_f(X)
\not= \emptyset.$ The proof is complete.

\

{\it Remark 4}.~~The exhaustion argument in the above proof goes
back to Yan [24].

\section*{Acknowledgment}

This paper was partly written when the author visited Equipe de
Math\'{e}matiques, Universit\'{e} de Franche-Comt\'{e}, supported
by the grant of the academic cooperation agreement between CAS and
CNRS. He would like to thank the Equipe de Math\'{e}matiques for
its kind hospitality. In particular, he is very grateful to
Professor Q.Xu for conversations on non-commutative probability.

\begin{center}R{\small EFERENCES}\end{center}
\begin{description}
\item[~1]~L.Bachelier, Th\'{e}orie de la sp\'{e}culation, {\it Ann. Sci.
\'{E}cole Norm. Sup.,} 17, 21-86 (1900). English translation in:
{\it The Random Character of Stock Market Prices,} P.Cootner ed.,
MIT Press, Cambridge, Mass., 1964, pp 17-78
\item[~2]~F.Black, M.Scholes, The pricing of options and corporate
liabilities, {\it J. Political Econ.,} 81, 637-659 (1973)
\item[~3]~Zeqian Chen, The meaning of quantum finance, submitted to
{\it Acta Math.Sci.}(Chinese Edition)
\item[~4]~Zeqian Chen, Quantum Finance: The Finite Dimensional
Cases, (quant-ph/0112158v2), 2001
\item[~5]~Zeqian Chen, Quantum theory for the binomial model in
finance theory, (quant-ph/0112156v5)
\item[~6]~R.C.Dalang, A.Morton, W.Willinger, Equivalent martingale
measures and no-arbitrage in stochastic securities market model,
{\it Stochastics Stoch. Rep.,} 29, 185-201 (1990)
\item[~7]~F.Delbaen, W.Schachermayer, A general version of
the fundamental theorem of asset pricing, {\it Math. Ann.,} 300(1994),
463-520
\item[~8]~F.Delbaen, W.Schachermayer, The fundamental theorem of asset
pricing for unbounded stochastic processes, {\it Math. Ann.,} 312(1998),
215-250
\item[~9]~F.Delbaen, W.Schachermayer, Applications to mathematical
finance, In: {\it Handbook of the Geometry of Banach Spaces,} W.B.Johnson
and J.Lindenstrauss, eds, Elsevier Science B.V., Amsterdam, 2001, 367-391
\item[10]~D.Duffie, C.F.Huang, Multiperiod security markets with
differential information: martingales and resolution times, {\it
J. Math. Econom.,} 15, 283-303 (1986)
\item[11]~J.M.Harrison, D.M.Kreps, Martingales and arbitrage in
multiperiod securities markets, {\it J. Econ. Theory,} 20(1979),
381-408
\item[12]~J.M.Harrison, S.R.Pliska, Martingales and stochastic
interfrals in the theory of continuous trading, {\it Stoch. Proc. Appl.,}
11(1981), 215-260
\item[13]~J.C.Hull, {\it Options, Futures and Other
Derivatives,} 4th Edition, Prentice-Hall, Inc., Prentice Hall,
2000
\item[14]~D.M.Kreps, Arbitrage and equilibrium in economies with
infinitely many commodities, {\it J. Math. Econ.,} 8(1981), 15-35
\item[15]~R.C.Merton, Theory of rational option pricing, {\it Bell J.
Econ. Manag. Sci.,} 4, 141-183 (1973)
\item[16]~R.C.Merton, {\it Continuous-time Finance,} Black-Well, Basil,
1990
\item[17]~G.Pisier, Q.Xu, Non-commutative martingale
inequalities, {\it Comm. Math. Physics,} 189, 667-698 (1997)
\item[18]~S.Sakai, {\it $C^*$-Algebras and $W^*$-Algebras,}
Springer-Verlag, Berlin, 1971
\item[19]~H.H.Sch\"{a}fer, {\it Topological Vector Spaces,} Springer-
Verlag, Berlin, 1966
\item[20]~A.N.Shiryaev, {\it Essentials of Stochastic Finance $(Facts,
Models, Theory),$} World Scientific, Singapore, 1999
\item[21]~C.Stricker, Arbitrage et lois de martingale, {\it Ann. Inst.
H. Poincar\'{e}, Probab. Statist.,} 26(1990), 451-460
\item[22]~M.Takesaki, Conditional expectations in von Neumann
algebras, {\it J. Funct. Anal.,} 9(1972), 306-321
\item[23]~M.Takesaki, {\it Theory of Operator Algebras I,}
Springer, New York, 2001
\item[24]~Jia'an Yan, Caracterisation d'une classe d'ensembles convexes
de $L^1$ ou $H^1,$ {\it Lecture Notes in Math.,} vol.784,
Springer-Verlag, 1980, 220-222
\end{description}
\end{document}